\documentclass[prl,showpacs,twocolumn,preprintnumbers,amsmath,amssymb]{revtex4}

\usepackage{epsfig}
\usepackage{graphicx}
\usepackage{dcolumn}
\usepackage{bm}

\begin{document}

\title{Gauge fields, ripples and wrinkles in graphene layers.}
\author{F. Guinea{$^1$}, Baruch Horovitz{$^2$} and P. Le Doussal{$^3$} }

{\affiliation{\affiliation{$^1$} Instituto de Ciencia de Materiales
de Madrid. CSIC. Cantoblanco. E-28049 Madrid, Spain}
\affiliation{{$^2$} Department of Physics, Ben Gurion University,
Beer Sheva 84105 Israel}
 \affiliation{{$^3$} CNRS-Laboratoire de
Physique Th{\'e}orique de l'Ecole Normale Sup{\'e}rieure, 24 rue
Lhomond,75231 Cedex 05, Paris France.}

\begin{abstract}
We analyze elastic deformations of graphene sheets which lead to
effective gauge fields acting on the charge carriers. Corrugations
in the substrate induce stresses, which, in turn, can give rise to
mechanical instabilities and the formation of wrinkles. Similar
effects may take place in suspended graphene samples under tension.
\end{abstract}

\maketitle
{\em Introduction.}
Two dimensional graphene membranes define a new class of materials
with a number of novel and promising
features\cite{Netal04,Netal05,NGPNG08}. From a basic point of view,
one of the striking properties of graphene is that many types of
long wavelength disorder influence the electrons as effective gauge
fields\cite{GGV92,GGV93,GGV01,Metal06,MG06,KN07b,CV07,JCV07,GKV08,GHL08,Wetal08,PN08},
which modify the electronic properties, as measured, for instance,
in weak localization experiments\cite{Metal06,Metal06b,MG06,THGS08}.

Because of the existence of two inequivalent valleys in the
Brillouin Zone of graphene, one can define formally two types of
gauge fields, one of which induces mixing between the valleys, and
another which does not. These two fields do not commute. A gauge
field which effectively rotates electrons from one valley into
electrons in the other valley is  needed to describe lattice
disclinations, such as pentagonal and heptagonal
rings\cite{GGV92,GGV93,LC04,KO06}. Intravalley gauge fields can be
induced by elastic strains\cite{SA02b,Metal06,M07,GKV08}, by
curvature\cite{KN07b}, or  by lattice defects such as dislocations.
For each type of field, the hamiltonian can be written as two
independent Dirac equations, with gauge fields of opposite sign for
each of them, so that it is invariant with respect to time reversal.

Elastic strains of reasonable values, $\bar{u} \sim 0.02 - 0.05$
with the correct symmetries can lead to significant effective
magnetic fields, $B \sim 0.1 - 1$T\cite{Metal06,GKV08}. Large
graphene sheets show corrugations\cite{Metal07}, and significant
scale deformations\cite{Betal08,Getal08,Betal08b}. Suspended
samples\cite{DSBA08,Betal08b} can also be deflected if a gate
potential is applied to them\cite{FGK08}. Graphene deposited on a
substrate shows corrugations\cite{ICCFW07,Setal08,Getal08,LLA08}.
The existence of corrugations is confirmed by numerical
simulations\cite{Fetal07}.

The low energy electronic states of graphene are described by the
Dirac equation\cite{NGPNG08}. The properties of the electronic
spectrum of the Dirac equation in the presence of a random gauge
field has been studied in connection with the Quantum Hall
Effect\cite{LFSG94,RH01,HD02}. The scattering cross section of
isolated ripples vanishes at short and long electron wavelengths,
with a peak for wavelengths comparable to the size of the
ripple\cite{G08}. The density of states is changed at low energies,
and it acquires an anomalous power law dependence on energy, where
the exponent depends on the strength of the disorder. The combined
effect of random gauge fields and the long range electron-electron
interaction leads to a complex phase diagram when the system is
close to the neutrality point\cite{SGV05,FA08,HJV08}.

In the following, we discuss some basic features of ripples and
wrinkles in single layer graphene. We analyze in the following
section the elastic strains in a periodic array of ripples, similar
to that reported in\cite{Vetal08}, and the effect of the induced
gauge fields. We next discuss possible mechanisms which may induce
wrinkles\cite{CM03} in graphene, and the way they can alter the
electronic properties.

We will not consider here the effect of a scalar potential also
induced by the presence of strains\cite{SA02b}. When the length
scale of the corrugations is larger than the Thomas-Fermi screening
length $k_s^{-1} \sim k_F^{-1}$, where $k_F$ is the Fermi
wavevector, these scalar fields will be screened\cite{FGK08}, while
the gauge fields are not. This picture is confirmed by full
electronic calculations of rippled graphene layers\cite{Wetal08}.

{\em Periodic array of ripples.} The simplest corrugation which
leads to inhomogeneous strains and, as a result, to non trivial
gauge fields in graphene is a periodic array of ripples, which can
be induced by a suitable substrate\cite{Vetal08}. For simplicity, we
assume the existence of modulation of the out of plane displacement
of the graphene layer of the type:
\begin{equation}
h ( x , y ) = h_0 \left[ 2 \cos ( \vec{G}_1 \vec{r} ) + 2 \cos (
\vec{G}_2 \vec{r} ) + 2 \cos ( \vec{G}_3 \vec{r} )\right]
\label{height}
\end{equation}
where $\vec{G}_1 = G {\bf n_x} , \vec{G}_2 = G ( {\bf n_x} / 2 +
\sqrt{3} {\bf n_y} / 2 )$ and $\vec{G}_3 = G ( - {\bf n_x}/2 +
\sqrt{3} {\bf n_y}/2 )$. These vectors define a triangular lattice.
In terms of the in-plane displacements, $\vec{u} ( x , y )$, the
strain tensor is:
\begin{align}
u_{xx} &= \partial_x u_x + \frac{\left( \partial_x h \right)^2}{2}
\nonumber \\
u_{yy} &= \partial_y u_y + \frac{\left( \partial_y h \right)^2}{2}
\nonumber \\
u_{xy} &= \frac{\partial_y u_x + \partial_x u_y}{2} + \frac{\left(
\partial_x h \right) \left( \partial_y h \right)}{2}
\end{align}
Neglecting for the moment the contribution  due to the bending
stiffness of the graphene layer, the free energy can be written
as\cite{LL59}:
\begin{equation}
{\cal F} \equiv \frac{\lambda}{2} \int d^2 \vec{r} \left( u_{xx} +
u_{yy} \right)^2 + \mu \int d^2 \vec{r} \left( u_{xx}^2 + u_{yy}^2 +
2 u_{xy}^2 \right) \label{free}
\end{equation}
where, using for the longitudinal and transverse sound velocities of
graphene\cite{NWS72,BCP88,BKMMT07} the values $v_L = 22.2$ Km/s and
$v_T = 14.7$ Km/s, we find $\lambda = 1.6$eV \AA$^{-2}$ and $\mu =
5.7$eV \AA$^{-2}$.

\begin{figure}
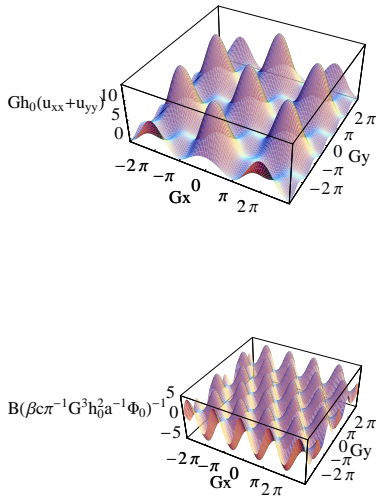

\begin{center}
\includegraphics*[width=5cm]{compression_periodic.dat}
\includegraphics*[width=5cm]{bmag_periodic.dat}
\end{center}
\caption{(Color online). Top: Compression, $u_{xx}+u_{yy}$, induced
by a periodic array of ripples, see  text for details. Bottom:
Effective magnetic field for the same array.}\label{strain_periodic}
\end{figure}

The full strain tensor $u_{ij}$ can be obtained from the tensor
$f_{ij} = \left( \partial_i h \right) \left( \partial_j h \right)$
by minimizing the free energy, eq.~(\ref{free}), and obtaining the
in plane displacement field, $\vec{u}$\cite{GHL08}. In terms of the
strain tensor, the effective gauge field acting on the electrons,
for a given valley in the Brillouin Zone, is\cite{SA02b,M07}:
\begin{align}
A_x ( \vec{r} ) &= \frac{c \Phi_0}{a} \frac{\partial \log ( t
)}{\partial \log ( a )} \left( u_{xx} - u_{yy} \right) \nonumber \\
A_y ( \vec{r} )&= 2 \frac{c \Phi_0}{a}
 \frac{\partial \log ( t )}{\partial \log ( a )} u_{xy} \label{gauge}
\end{align}
where $\beta = \partial \log ( t ) / \partial \log ( a ) \approx 2$
gives the dependence of the tight binding hopping element between
orbitals in nearest neighbor carbon atoms, $t \approx 3$eV, on the
distance between them, $a \approx 1.4$\AA, $c$ is a numerical
constant of order unity, and $\Phi_0$ is the quantum unit of flux.
In this analysis, we have assumed that the shape of the corrugation,
$h ( x , y )$, is determined by external forces. We now consider the
bending rigidity of the layer. From the continuum theory of
elasticity\cite{LL59} we expect a term:
\begin{align}
{\cal F}_b &= \frac{\kappa_1}{2} \int d^2 \vec{r} \left[ \left(
\frac{\partial^2 h}{\partial x^2} \right) + \left( \frac{\partial^2
h}{\partial y^2} \right) \right]^2 + \nonumber \\
&+ \frac{\kappa_2}{2} \int d^2 \vec{r} \left[ \left(
\frac{\partial^2 h}{\partial x^2} \right)^2 + \left(
\frac{\partial^2 h}{\partial y^2} \right)^2 + \frac{1}{2} \left(
\frac{\partial^2 h}{\partial x \partial y} \right)^2 \right]
\label{bending}
\end{align}
For graphene, $\kappa_1 \sim \kappa_2 \sim \bar{\kappa} \sim 1$eV.
The ratio between the bending and compression energies of the
ripples is approximately proportional to $G^2 \bar{\kappa} / (
\lambda + 2 \mu )$. For mesoscopic ripples of lengths $l = 2 \pi
G^{-1} \sim 10 - 100$nm, the bending rigidity can be neglected.

\begin{figure}
\begin{center}
\includegraphics*[width=5cm]{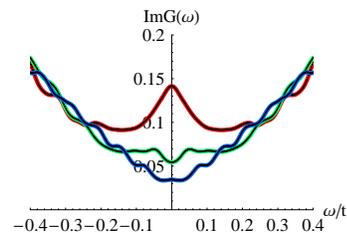}
\end{center}
\caption{(Color online). Density of states for a periodic array of
ripples, including the induced gauge field. Blue: $h_0 = 0$.
 Green: $\frac{3 c \beta a G}{2} ( G h_0 )^2 = 0.025$.
Red: $\frac{3 c \beta a G}{2} ( G h_0 )^2 = 0.05$
}\label{dos_periodic}
\end{figure}
The calculation of the strain tensor and effective gauge fields for
the height profile in eq.~(\ref{height}) is rather simple. Values
for the strain and induced magnetic field are shown in
Fig.~\ref{strain_periodic}. The electronic density of states,
calculated including the gauge field due to the ripples, is shown in
Fig.~\ref{dos_periodic}. The results have been obtained by using a
discrete lattice model with $2 \times 81 \times 81$ sites with a
broadening of the levels of $\delta \omega = 0.03 t$. The small
amplitude undulations in the density of states are an artifact due
to the discretization. There is a peak at the Dirac energy for the
largest corrugations, consistent with analytical results for the
density of states of Dirac electrons in the presence of a random
gauge field\cite{HD02}. Similar effects are obtained in numerical
studies of the Dirac equation in a random field whose fluctuations
are smooth on the scale of the lattice\cite{unpub}, although they
have not be found if the gauge field varies on length scales
comparable to the lattice spacing\cite{RH01}. The highest effective
magnetic fields induced by ripples with $h_0 \sim 1$nm and $l = 2
\pi G^{-1} \sim 60$nm can be large, $B_{max} \sim
2-5$T\cite{Metal06}, so that the number of flux quanta per ripple
exceeds one.

\begin{figure}
\begin{center}
\includegraphics*[width=5cm]{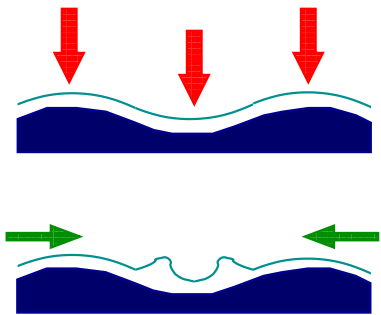}
\end{center}
\caption{(Color online). Generation of wrinkles in a graphene layer
on SiO$_2$. Top: The layer is compressed against the corrugated
substrate. Bottom: When the external pressure is removed, in-plane
stresses due to the deformation of the layer induce wrinkles, if the
pinning to the substrate is sufficiently
weak.}\label{sketch_crumpling}
\end{figure}
{\em Wrinkling of graphene layers: Graphene on a substrate.} As in
any membrane, topological lattice defects can spontaneously lead to
corrugations\cite{SN88} in graphene (see also Sec. VIB
in~\cite{GHL08}). Similar effects can be expected from impurities
which induce lattice deformations\cite{TMM08}. Rigid layers on soft
substrates show wrinkles if the are under anisotropic stresses which
exceed a given threshold\cite{CM03,GG06,W07}.  A related situation
which may be relevant for graphene layers on SiO$_2$ substrates is
sketched in Fig.~\ref{sketch_crumpling}. The exfoliation procedure
applies pressure on the graphene layer against the corrugated
substrate, inducing deformations and in plane stresses. When the
pressure is removed, the system can reduce the elastic energy by
forming wrinkles, provided that the stresses are sufficiently large
and the pinning energy to the substrate is sufficiently low.

\begin{figure}
\begin{center}
\includegraphics*[width=5cm]{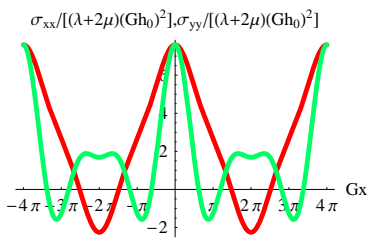}
\end{center}
\caption{(Color online). Stresses along the line $(x,0)$ for the
periodic array of ripples analyzed in Fig.~\ref{strain_periodic}.
Red: $\sigma_{xx} ( x,0)$. Green: $\sigma_{yy}
(x,0)$.}\label{stress}
\end{figure}
We show in Fig.~\ref{stress} the stresses $\sigma_{xx}$ and
$\sigma_{yy}$ induced by the periodic array of ripples discussed in
the previous section. There are regions, whose length is fraction of
$G^{-1}$, where the stresses are highly anisotropic. The gain in
elastic free energy per unit area due to the formation of wrinkles
in these regions is ${\cal F} \sim ( \lambda + 2 \mu ) ( G h_0 )^2
\sim 0.05 - 0.1$eV \AA$^{-2}$, for $h_0 \sim 1$nm and $l = 2 \pi
G^{-1} \sim 60$nm. This value is larger than typical interactions
energies between a graphene layer and  a SiO$_2$ substrate, or other
materials, like water or ions, which may be trapped between graphene
and the substrate\cite{SSFGNS08}. Hence, we expect that, if graphene
is pressed onto a corrugated SiO$_2$ substrate, the removal of that
pressure will lead to the spontaneous formation of wrinkles. Note
that the lowering of strains will actually reduce the effective
gauge field acting on the electrons. On the other hand, the regions
detached from the substrate can support low energy flexural modes,
which can scatter electrons at finite temperatures\cite{MO08,KG08}.

{\em Wrinkling of graphene layers: suspended graphene samples}. We
now analyze the structural instabilities towards curved shapes,
wrinkles, which can  arise in suspended membranes under
tension\cite{GDLP88}, see also Sec. VII in~\cite{GHL08}. A stretched
membrane with clamped ends will develop wrinkles if the applied
tension is large enough\cite{CM03}. This threshold arises from a
balance between the bending energy and the applied
tension\cite{CM03}. Wrinkles will occur when the tension, $T \sim (
\lambda + 2 \mu ) u_{xx}$ exceeds $\bar{\kappa} / l^2$, where $l$ is
the length of the stretched region. Hence,  in graphene, stresses
such that $u_{xx} \sim 10^{-2}$, will lead to wrinkles when $l
\gtrsim 1$nm. The resulting deformation is periodic, with a
wavelength, $l_w$, and amplitude, $A$, in the direction
perpendicular to the applied strain given by\cite{CM03}:
\begin{align}
l_w &= 2 \sqrt{\pi} \left( \frac{\bar{\kappa}}{T} \right)^{1/4}
l^{1/2} \nonumber \\
A &=\frac{\sqrt{2}}{\pi} \left(\frac{\lambda \bar{u}}{\lambda+2\mu}
\right)^{1/2} l_w \label{wrinkles}
\end{align}

\begin{figure}
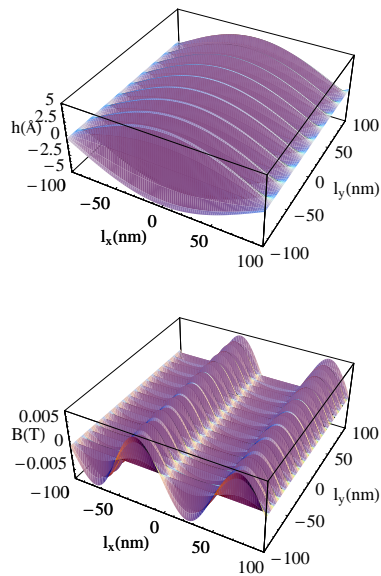

\begin{center}
\includegraphics*[width=5cm]{wrinkles.dat}
\includegraphics*[width=5cm]{bmag_wrinkles.dat}
\end{center}
\caption{(Color online). Top: Wrinkles expected in a suspended
graphene sheet of $l = 200$nm under an applied tension $\Delta l / l
= 2 \times 10^{-2}$. Bottom: Effective magnetic field induced by the
wrinkles shown in the top.}\label{wrinkles_fig}
\end{figure}
Typical wrinkles, calculated using eq.~(\ref{wrinkles}) for a
suspended region of length $l = 200$nm, and an applied strain
$\bar{u} = 2 \times 10^{-2}$ are shown in Fig.~\ref{wrinkles_fig}.

A suspended sheet with constant tension gives rise to a constant
gauge field\cite{FGK08}. Electron scattering takes place at the
boundaries between the region with and without tension, and the the
carriers propagate freely within the suspended region. Wrinkles
modulate the strains  in the direction of the applied tension and
also in the direction perpendicular to it. As a result, an effective
magnetic field is induced throughout the suspended region. This
field is also shown in Fig.~\ref{wrinkles_fig}.  Its average value
scales as:
\begin{equation}
\bar{B} \sim \frac{c \beta A^2 \Phi_0}{l_w^2 l a } \sim  \frac{c
\beta \bar{u} \Phi_0}{ l a }
\end{equation}
Electronic transport in suspended graphene samples is expected to be
close to the ballistic limit\cite{Betal08a}. Hence, even the low
fields induced by wrinkles may lead to observable effects.

{\em Effects of effective gauge fields on Landau levels.} The gauge
fields induced by stresses modify the Landau levels induced by a
magnetic field, and give to them an index dependent width, see Sec.
VIII in~\cite{GHL08}. The energy of the n-th Landau level scales as
$\epsilon_n  \propto \sqrt{nB}$ where $B$ is the magnetic field.
Stresses modify the effective field so that, to a first
approximation the width of the Landau level becomes proportional to
$\sqrt{n}$. This effect is partially compensated by the averaging of
the effective field within the spatial extent of the Landau level,
given by a scale $\propto \sqrt{n} l_B$, where $l_B = \sqrt{\Phi_0 /
( \pi B )}$ is the magnetic length, and $\Phi_0$ is the quantum unit
of flux. The narrowest Landau level is that with $n=0$, in agreement
with experiments\cite{Getal07}.

 {\em
Conclusions.} We have analyzed possible deformations of graphene
layers, and the effective gauge fields that they induce. The shape
of the layer can be fixed by the corrugated substrate to which it is
pinned. If the deformations induce substantial stresses, wrinkles
can be produced spontaneously. A similar elastic instability leads
to wrinkles in suspended graphene samples.

{\em Acknowledgements.}
 This work was supported by MEC (Spain)
through grant FIS2005-05478-C02-01 and CONSOLIDER CSD2007-00010, by
the Comunidad de Madrid, through CITECNOMIK, CM2006-S-0505-ESP-0337,
and the EU Contract 12881 (NEST).
\bibliography{bib_suspended_2}
\end{document}